\documentclass[lettersize,journal]{IEEEtran}
\usepackage{amsmath,amsfonts}
\usepackage{algorithmic}
\usepackage{algorithm}
\usepackage{array}
\usepackage[caption=false,font=normalsize,labelfont=sf,textfont=sf]{subfig}
\usepackage{textcomp}
\usepackage{stfloats}
\usepackage{url}
\usepackage{multirow}
\usepackage{verbatim}
\usepackage{graphicx}
\usepackage{cite}
\usepackage{booktabs}
\begin{document}

\title{Towards Multimodal Sentiment Analysis via Contrastive Cross-modal Retrieval Augmentation and Hierachical Prompts}

\author{Xianbing Zhao, Shengzun Yang, Buzhou Tang, Ronghuan Jiang
\thanks{
Xianbing Zhao is with School of Artificial Intelligence and Computer Science,
Jiangnan University, Wuxi, China (e-mail: zhaoxianbing\_hitsz@163.com).
}
\thanks{
Shengzun Yang, Xianbing Zhao, Buzhou Tang are with School of Computer Science and Technology, Harbin Institute of Technology, Shenzhen, Guangdong, 518055, China (e-mail: 24S151174@stu.hit.edu.cn and tangbuzhou@hit.edu.cn).
}

\thanks{Ronghuan Jiang is with Chinese People’s Liberation Army General Hospital, Beijing, China. (e-mail: jangrh55@126.com).}

\thanks{Buzhou Tang and Ronghuan Jiang are co-corresponding authors.}
}

\markboth{Journal of \LaTeX\ Class Files,~Vol.~14, No.~8, August~2021}%
{Shell \MakeLowercase{\textit{et al.}}: A Sample Article Using IEEEtran.cls for IEEE Journals}


\maketitle

\begin{abstract}
Multimodal sentiment analysis is a fundamental problem in the field of affective computing. Although significant progress has been made in cross-modal interaction, it remains a challenge due to the insufficient reference context in cross-modal interactions. Current cross-modal approaches primarily focus on leveraging modality-level reference context within a individual sample for cross-modal feature enhancement, neglecting the potential cross-sample relationships that can serve as sample-level reference context to enhance the cross-modal features. To address this issue, we propose a novel multimodal retrieval-augmented framework to simultaneously incorporate inter-sample modality-level reference context and cross-sample sample-level reference context to enhance the multimodal features. In particular, we first design a contrastive cross-modal retrieval module to retrieve semantic similar samples and enhance target modality. To endow the model to capture both inter-sample and intra-sample information, we integrate two different types of prompts, modality-level prompts and sample-level prompts, to generate modality-level and sample-level reference contexts, respectively. Finally, we design a cross-modal retrieval-augmented encoder that simultaneously leverages modality-level and sample-level reference contexts to enhance the target modality.
    Extensive experiments demonstrate the effectiveness and superiority of our model on two publicly available datasets.

\end{abstract}


\maketitle

\section{Introduction}
    \IEEEPARstart{M}ultimodal content, primarily consisting of text, visual, and acoustic, is widely present in our daily lives. It is crucial for computers to analyze and understand these multimodal data. Multimodal sentiment analysis is an important and fundamental problem in the field of affective computing, which has attracted significant attention from researchers and made substantial progress.  Previous multimodal sentiment analysis studies can be broadly divided into two categories: multimodal interaction learning methods and multimodal representation learning methods. The former learns enhanced multimodal features through cross-modal interaction learning \cite{tsai2019multimodal,rahman2020integrating,yang2021mtag,lv2021progressive,han2021bi}, while the latter learns fine-grained representations by modeling the geometric relationships in the cross-modal representation space \cite{yu2021learning,yang2022disentangled,zhao2023tmmda,li2023decoupled,yang2023confede,zeng2024disentanglement}. These approach effectively addresses key limitations of multimodal interaction learning, such as the alignment and fusion of cross-modal information. 

\begin{figure}
    \centering
    \includegraphics[width=\linewidth]{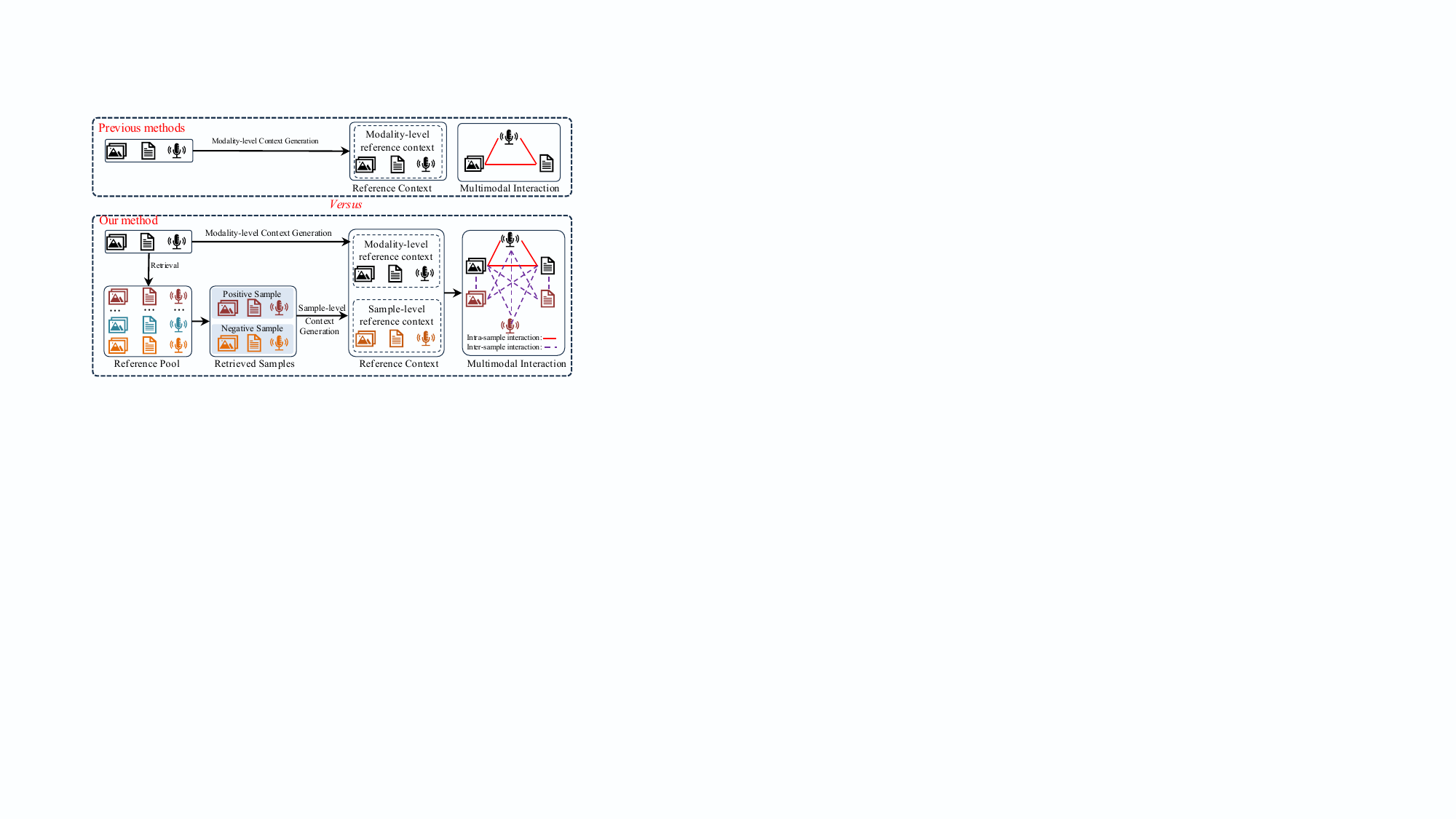}
    \caption{
    Previous cross-modal interaction paradigm that solely rely on modality-level reference context to enhance target modality produce erroneous sentiment polarity. In contrast, our method simultaneously utilizes both modality-level and sample-level reference contexts to enhance target modality, resulting in the correct sentiment polarity.
    }
    \label{fig:premodel}
\end{figure}

Despite achieving significant progress, these methods still suffer from important drawbacks. These approaches learn enhanced multimodal features that are limited to the information within individual sample, overlooking the rich reference context information provided by similar samples. The reference context offered by similar samples can provide more richer information for multimodal feature enhancement compared to intra-sample multimodal interaction learning \cite{gao2023retrieval,long2022retrieval,ram2023context,blattmann2022retrieval,gur2021cross}.  As illustrated in Figure \ref{fig:premodel}, the multimodal sentiment analysis model that only considers modality-level reference context within the individual sample generates incorrect sentiment polarity. By additionally incorporating sample-level reference context of intra-samples, the multimodal sentiment analysis model produces the correct sentiment polarity. Therefore, it would be interesting to investigate i) how to introduce cross-sample reference context into multimodal interaction learning, ii) how to efficiently facilitate the integration of inter-sample and intra-sample reference contexts to learn enhanced multimodal features.

To tackle these downsides, we propose a novel \textbf{T}owards Multimodal Sentiment Analysis via \textbf{C}ontrastive \textbf{C}ross-modal \textbf{R}etrieval \textbf{A}ugmentation and \textbf{H}ierarchical \textbf{P}rompts, namely \textbf{TC$^2$RAHP}, which is the first unified multimodal sentiment analysis framework that can introduce and incorporate modality-level and sample-level reference context to enhance target modality. As shown in Figure \ref{fig:model}, we first design a contrastive cross-modal retrieval module to learn the reference context of similar samples. This module learns positive and negative samples similar to the target modality under the guidance of positive and negative sentiment semantics. We leverage contrastive learning to constrain the model to learn samples that are semantically similar to the target modality, rather than just feature-wise similarity. Positive samples are selected as sample-level reference contexts to enhance the target modality. To facilitate inter-sample and intra-sample interaction learning, we design a prompt-based cross-modal reference context generation module. This module includes two different types of prompts, modality-level prompts and sample-level prompts, which map cross-modal and cross-sample information into modality-level and sample-level reference context. We design a cross-modal retrieval-enhanced encoder that leverages modality-level and sample-level reference contexts to enhance the target modality and improve its representational ability, thereby increasing the accuracy of sentiment polarity prediction. Extensive experiments on two publicly available multimodal datasets, CMU-MOSI and CMU-MOSEI, validate the effectiveness and superiority of our model. The main contributions of this work are threefold:
\begin{itemize}
    \item We introduce a cross-modal retrieval-augmented interaction framework for multimodal sentiment analysis, which is capable of learning cross-modal and cross-sample information to enhance the target modality. 
    \item We design a contrastive cross-modal retrieval module that retrieves positive and negative samples relative to the target modality and strengthens the semantics of these samples.
    \item We design a prompt-based cross-modal reference context generation module that maps inter-sample and intra-sample information into modality-level and sample-level reference contexts. Additionally, we customize a cross-modal retrieval-augmented encoder that utilizes these reference contexts to enhance the target modality.
\end{itemize}

\section{Related Work}

\subsection{Multimodal Sentiment Analysis}
Multimodal Sentiment Analysis refers to the process of comprehensively analyzing and understanding human sentiment by leveraging multiple modalities \cite{han2021improving,tsai2019learning}. Current research focuses primarily on two aspects: multimodal representation learning and multimodal fusion.

The former continuously explores semantics \cite{chen2017multimodal,zong2023acformer} through feature transformation to obtain more fine-grained feature representations. Zadeh et al. \cite{zadeh2017tensor} further explored the local information of the features by reshaping and concatenating the tensors, as well as establishing associations between the different tensors. Liu et al. \cite{liu2018efficient} deconstructed the matrix to transform it into a low-rank matrix, capturing feature information in the altered vector space. Yang et al. \cite{yang2022disentangled} decomposed the features into fine-grained shared and private features through adversarial methods, capturing information along these two dimensions. Yu et al. \cite{yu2021learning} synthesized global information to generate sentiment labels, which then guided further feature extraction. The latter projects multiple non-aligned local features into the same feature space through interactions between different modalities and obtains global sentiment information within this feature space \cite{mai2023excavating}. Tsai et al. \cite{tsai2019multimodal} first projected multiple local features into the same vector space through cross-modal attention and then integrated global sentiment information through self-attention. Han et al. \cite{han2021improving} designed a hierarchical structure to layer-wise integrate information from various modalities, as well as to fuse information across modalities and within individual modalities, for sentiment analysis. Yu et al. \cite{yu2023conki} assigned modality-specific information to each modality to guide the extraction of features and the fusion of global information. Zhang et al. \cite{zhang2025modal} identified the modules with weak information extraction capabilities and prioritized training these modules through targeted prompts. Li et al. \cite{li2025t} noticed the noise caused by modal imbalance and mitigated it by adjusting the distance between modalities in the semantic space to perform feature denoising. 

With the development of LLMs and MLLMs \cite{liang2024survey}, many studies have begun to explore the application of LLMs and MLLMs in multimodal sentiment analysis tasks \cite{mixtralai,gemmagoogle}, demonstrating superior zero-shot performance. LLMs such as ChatGPT \cite{chatgpt2023} and LLama \cite{touvron2023llama} have shown excellent sentiment analysis results by capturing richer feature representations in text. MLLMs such as Gemini-V \cite{geminiv} and BLIP-2 \cite{li2023blip2bootstrappinglanguageimagepretraining} possess stronger cross-modal interaction capabilities, allowing better utilization of information from different modalities.

\subsection{Retrieval Augmentation}
Retrieval Augmentation refers to the process of incorporating relevant external information into the current data by retrieval, thus improving the representation capacity of existing data \cite{wang2020joint,gur2021cross,blattmann2022retrieval,gao2023retrieval,jiang2023active}. Long et al. \cite{long2022retrieval} added an image library and a retrieval module to a basic image feature extraction module, improving image classification capabilities by introducing additional image information. Xu et al. \cite{xu2023weakly} improved the performance of the document classifier by mapping the query data and the retrieved data to the same feature space for matching. Ram et al. \cite{ram2023context} enhanced the semantics of the text by incorporating the retrieved textual information into the input while keeping the model structure fixed. Lewis et al. \cite{lewis2020retrieval} combined retrieval augmentation techniques with large models and provided a widely applicable fine-tuning framework, achieving excellent results. Rezaei et al. \cite{rezaei2025vendi} designed an evaluation score to balance the richness and accuracy of generated content during multi-round retrieval. Yang et al. \cite{yang2025timerag} constructed a knowledge base based on information from existing sequences, and utilized the retrieved knowledge to interact with the current sequence, thereby guiding predictions for future sequences.

\begin{figure*}
    \centering
    \includegraphics[width=\textwidth]{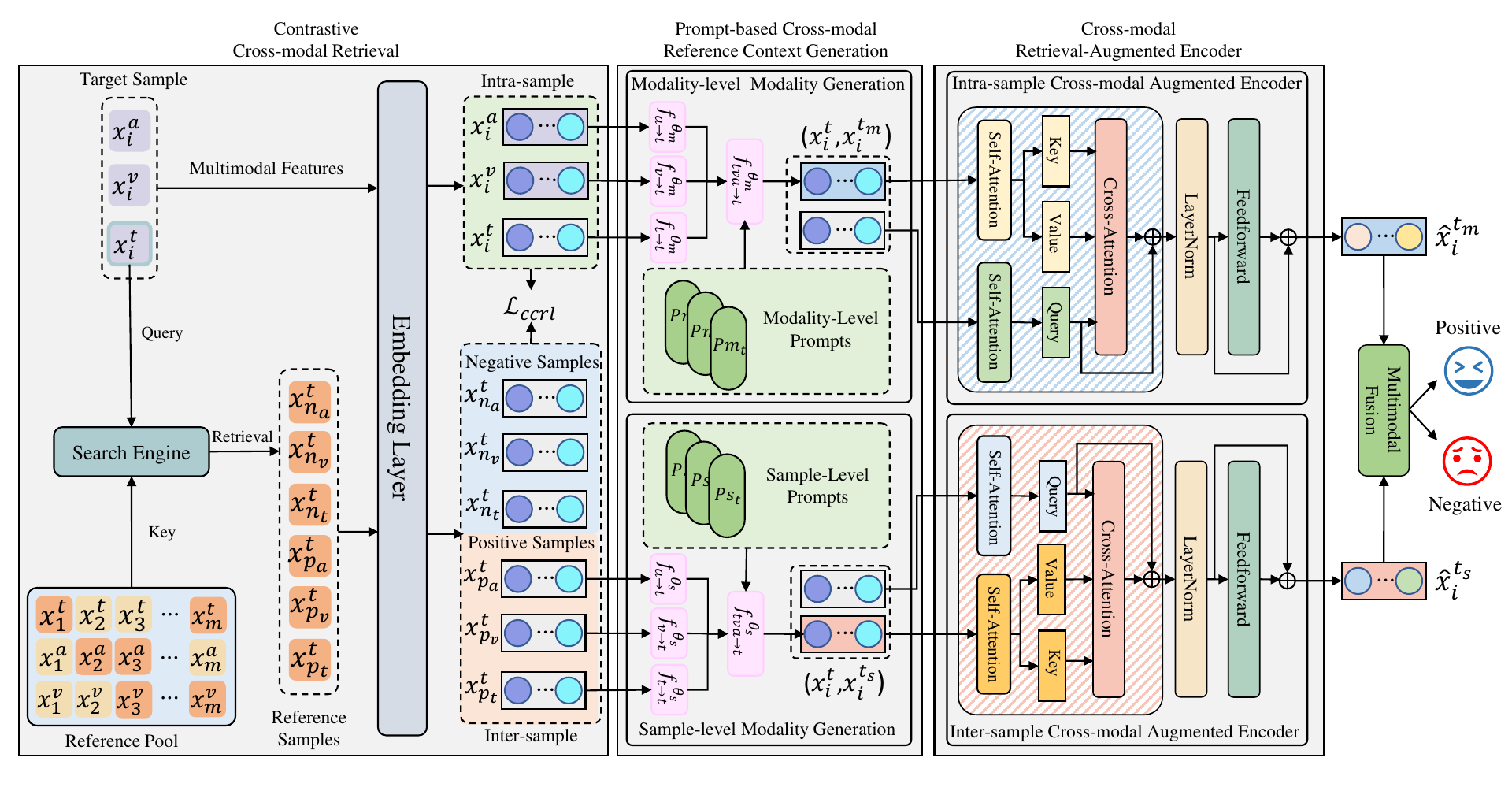}
    \caption{Schematic illustration of the proposed TC$^2$RAHP framework with three components, Cross-modal Contrastive Retrieval module, Prompt-based Cross-modal Generation, and Cross-modal Retrieval-Augmented Encoder. 
    }
    \label{fig:model}
\end{figure*}
\section{METHODOLOGY}

In this section, we customize three modules, as illustrated in Figure \ref{fig:model}. Specifically, we first introduce the contrastive cross-modal retrieval module in Section \ref{sec:ccr} to learn the reference context from external samples. Next, we present the prompt-based cross-modal reference context generation module in Section \ref{sec:pcg}, which utilizes modality-level prompts and sample-level prompts to generate mappings for intra-sample and inter-sample information. Finally, in Section \ref{sec:cre}, we design the cross-modal retrieval-augmented encoder, which integrates the generated intra-sample and inter-sample information into the target modality to learn enhanced multimodal features.
\subsection{Overall Framework}
Figure 2 illustrates the workflow of our model. Taking the target modality as the text modality as an example. The workflow of the proposed model is as follows: Given the $i$-th video clip, we use the pre-trained models to extract text \cite{raffel2020exploring}, visual \cite{cao2018vggface2}, and acoustic \cite{hsu2021hubert} features, denoted as $x_i^{\{t,v,a\}}\in\mathcal{D}$, respectively. $\mathcal{D}$ denotes the whole dataset. 
We obtain the feature set $\{x_j^{\{t,v,a\}}\}\in\mathcal{D},i\neq j$ as the reference pool. 
Under the guidance of sentiment labels, we retrieve the most similar positive and negative samples relative to the target modality $x_i^{t}$ from the reference pool, denoted as $\{x_{n_{t}}^{t},x_{p_{t}}^{t}\}$, $\{x_{n_{v}}^{t},x_{p_{v}}^{t}\}$ and $\{x_{n_{a}}^{t},x_{p_{a}}^{t}\}$. We use contrastive loss $\mathcal{L}_{ccrl}$ to constrain the target modality and the retrieved samples. In the prompt-based cross-modal generation network, modality-level prompts $P_{m_t}$ and intra-sample features $x_{i}^{\{t,v,a\}}$ are input into the modality generation network $f^{\theta_m}_{*}$ to obtain modality-level reference context ${x}_{i}^{t_m}$. Sample-level prompt $P_{s_t}$ and inter-sample features $x_{p_{\{t,v,a\}}}^{t}$ are input into the sample generation network $f^{\theta_s}_{*}$ to obtain sample-level reference context ${x}_{i}^{t_s}$. The modality-level and sample-level reference context information ${x}_{i}^{t_m}$ and ${x}_{i}^{t_s}$ are respectively input into the cross-modal retrieval-augmented encoder to enhance the target modality $x_{i}^{t}$, thereby obtaining the enhanced features $\hat{x}_{i}^{t_m}$ and $\hat{x}_{i}^{t_s}$. Following the same pipeline, we can also obtain ($\hat{x}_{i}^{v_m}$,$\hat{x}_{i}^{v_s}$) and ($\hat{x}_{i}^{a_m}$,$\hat{x}_{i}^{a_s}$). Finally, the enhanced multimodal features are fused to obtain $x_{final}$ for sentiment polarity prediction.Subsequent sections provide details of the proposed components.

\subsection{Contrastive Cross-modal Retrieval}
\label{sec:ccr}
Previous multimodal sentiment analysis methods \cite{tsai2019multimodal,zhang2025modal,li2025t} primarily explore intra-sample multimodal interaction learning to enhance the features of the target modality, neglecting the reference information provided by external samples. To this end, we propose a contrastive cross-modal retrieval module that selects reference samples from external samples and provides richer reference context for feature enhancement of the target modality. This module is capable of effectively retrieving valid reference samples from external samples and optimizing the semantic of these reference samples with the guidance of sentiment category.
Specifically, given a mini-batch containing $\mathcal{M}\in\mathcal{D}$ samples as reference pool $\{x_{j}^{\{t,v,a\}}\}\in\mathcal{M}$, we use cosine similarity as the similarity measure to retrieve the samples with the highest feature similarity. The samples with matching sentiment labels to the target modality $m_\alpha\in\{t,v,a\}$ are selected as positive samples $x_{p_{m_\beta}}^{\{t,v,a\}},m_\beta\in\{t,v,a\}$, while those with opposing sentiment labels to the target modality are selected as negative samples $x_{n_{m_\beta}}^{\{t,v,a\}},m_\beta\in\{t,v,a\}$. 
The cross-modal retrieval similarity is calculated as follows: 
\begin{gather}
    x_{p_{m_\beta}}^{m_\alpha} = \arg\max_{x_j \in \mathcal{M}, i\neq j} \mathcal{F}(x_i^{m_\alpha}, x_j^{m_\beta}), m_\alpha,m_\beta\in{\{t,v,a\}}, y_i=y_j, \\
        x_{n_{m_\beta}}^{m_\alpha} = \arg\max_{x_j \in \mathcal{M}, i\neq j} \mathcal{F}(x_i^{m_\alpha}, x_j^{m_\beta}), m_\alpha,m_\beta\in{\{t,v,a\}}, y_i\neq y_j,
    \label{eq:sim}
\end{gather}
where $x_i^{m_\alpha}$ and $x_j^{m_\beta}$ represent the features of the target modality and the reference sample, respectively. $\mathcal{F}$ represents the cosine similarity function. The $y_i$ and $y_j$ represent the sentiment label of samples $i$ and $j$ at training stage, respectively. Note that, during the testing phase, the model is already able to effectively distinguish the semantics of positive and negative samples, so label guidance is not required. We can retrieve the corresponding positive and negative reference sample pairs $\{x_{p_{m_\beta}}^{m_\alpha},x_{n_{m_\beta}}^{m_\alpha}\}$ for each modality according to the Eq. \ref{eq:sim}. Formally,
\begin{equation}
      x_{p_{m_\beta}}^{m_\alpha} = \{x^t_{p_{\{t,v,a\}}}, x^v_{p_{\{t,v,a\}}}, x^a_{p_{\{t,v,a\}}}\}
\label{eq:pos}
\end{equation}
\begin{equation}
      x_{n_{m_\beta}}^{m_\alpha} = \{x^t_{n_{\{t,v,a\}}}, x^v_{n_{\{t,v,a\}}}, x^a_{n_{\{t,v,a\}}}\}
\end{equation}
where the superscript $m_\alpha$ represents the target modality, the subscript  $m_\beta$ represents the retrieved modality, and \(p\) and \(n\) denote positive and negative samples, respectively. For example, $x^t_{p_v}$ represents the most similar visual modality retrieved from external positive samples with the text modality as the target modality. Only leveraging the similarity between the target modality and external samples makes it difficult to learn semantic associations, as the target modality may have high similarity with external samples that possess opposite sentiment polarities. We design a contrastive loss to reduce the distance between the target modality and the positive reference sample features, while increasing the distance between the target modality and the negative reference sample features. The contrastive cross-modal retrieval loss is formulated as follows:
\begin{equation}
   \mathcal{L}_{ccrl} = \sum_{i\in \mathcal{D}}\sum_{m_\alpha, m_\beta }  {d}(x_{i}^{m_\alpha},x_{p_{m_\beta}}^{m_\alpha})^2
   + max\{0,\gamma - {d}(x_{i}^{m_\alpha},x_{n_{m_\beta}}^{m_\alpha})^{2}\}
\end{equation}
where $m_\alpha, m_\beta  \in \{t,a,v\}$ denotes modality, ${d}$ represents the $l_2$ norm constraint between the two feature spaces, and $\gamma$ denotes the upper bound of the distance metric between the target modality and the negative reference sample. The contrastive constraint ensures that the retrieved samples are not only feature-wise similar, but more importantly, semantically similar in terms of sentiment.

\subsection{Prompt-based Cross-modal Reference Context Generation}
\label{sec:pcg}
Multimodal data often exhibits gaps in feature space distributions \cite{hazarika2020misa,yu2023conki}. Previous methods address this issue by designing bridging neural networks to perform modality-to-modality mapping \cite{zhao2023tmmda}. Specifically, in scenarios with missing modality \cite{guo2024multimodal}, these methods use prompts to bridge the gap and generate the missing modality, as prompts can learn cross-modal mapping information. Based on this observation, we propose using prompts to transform cross-modal reference samples into representations of the target modality's reference context. Modality-level prompts and sample-level prompts are used for generating modality-level and sample-level reference contexts, respectively.  We refer to these prompts as hierarchical prompts, which can be formalized as: 
\begin{gather}
    P_{m} = \{P_{m_t},P_{m_v},P_{m_a}\} \\
    P_{s} = \{P_{s_t},P_{s_v},P_{s_a}\}
\end{gather}
where $P_{m}$ and $P_{s}$ represent modality-level and sample-level prompts, respectively. The initialization \cite{he2015delving} form of the prompt is as follows:
\begin{equation}
P_{\{s,m\}} \sim \mathcal{U}\left( -\sqrt{\frac{b}{(1 + a^2) \cdot fan\_in}}, \ +\sqrt{\frac{b}{(1 + a^2) \cdot fan\_in}} \right)
\end{equation}
where $fan\_in$ represents the number of input neurons, which is the feature dimension of the prompt. The symbol $a$ represents the negative slope parameter with a value of 5, and $b$ is a constant with a value of 6. The symbol $\mathcal{U}$ represents the uniform distribution, indicating that weights are sampled symmetrically within the interval. We inject prompt $P_m$ and $P_s$ into functions $f_{(\cdot)}^{\theta_m}$ and $f_{(\cdot)}^{\theta_s}$ with parameters $\theta_m$ and $\theta_s$, respectively, to transform the reference samples into the corresponding modality-level reference context. This can be formalized as:
\begin{gather}
    x^{t_s} = f_{tva\rightarrow t}^{\theta_s}([P_{s_t},f^{\theta_s}_{t\rightarrow t}(x^t_{p_t}),f^{\theta_s}_{v\rightarrow t}(x^{t}_{p_v}),f^{\theta_s}_{a\rightarrow t}(x^{t}_{p_a})]) \\
    x^{v_s} = f_{tva\rightarrow v}^{\theta_s}([P_{s_v},f^{\theta_s}_{t\rightarrow v}(x^v_{p_t}),f^{\theta_s}_{v\rightarrow v}(x^{v}_{p_v}),f^{\theta_s}_{a\rightarrow v}(x^{v}_{p_a})]) \\
    x^{a_s} = f_{tva\rightarrow a}^{\theta_s}([P_{s_a},f^{\theta_s}_{t\rightarrow a}(x^a_{p_t}),f^{\theta_s}_{v\rightarrow a}(x^{a}_{p_v}),f^{\theta_s}_{a\rightarrow a}(x^{a}_{p_a})]) 
\end{gather}
where $x^{t_s}$, $x^{v_s}$ and $x^{a_s}$ represent the sample-level reference context, which is used to enhance the features of the target modality. The $x^t_{p_{\{t,v,a\}}}$, $x^v_{p_{\{t,v,a\}}}$, and $x^a_{p_{\{t,v,a\}}}$ represents the positive reference samples retrieved from external samples, as defined in Eq. \ref{eq:pos} of Section \ref{sec:ccr}. Similarly, we also transform the intra-sample reference modality into the sample-level reference context to enhance the features of the target modality. Formally, this can be represented as:
\begin{gather}
    x^{t_m} = f_{tva\rightarrow t}^{\theta_m}([P_{m_t},f^{\theta_m}_{t\rightarrow t}(x^t),f^{\theta_m}_{v\rightarrow t}(x^{v}),f^{\theta_m}_{a\rightarrow t}(x^{a})]) \\
    x^{v_m} = f_{tva\rightarrow v}^{\theta_m}([P_{m_v},f^{\theta_m}_{t\rightarrow v}(x^t),f^{\theta_m}_{v\rightarrow v}(x^{v}),f^{\theta_m}_{a\rightarrow v}(x^{a})]) \\
    x^{a_m} = f_{tva\rightarrow a}^{\theta_m}([P_{m_a},f^{\theta_m}_{t\rightarrow a}(x^t),f^{\theta_m}_{v\rightarrow a}(x^{v}),f^{\theta_m}_{a\rightarrow a}(x^{a})]) 
\end{gather}
where $x^{t_m}$, $x^{v_m}$, and $x^{a_m}$ are the modality-level reference contexts used to enhance the features of the target modality. The $x^t$, $x^v$, and $x^a$ represent the multimodal features of individual intra-sample that need to be enhanced target modalities, respectively. Since both intra-sample and inter-sample multimodal information need to be transformed into reference contexts to enhance the target modality, the prompt-based transformation function effectively facilitates the conversion of cross-modal information.

\subsection{Cross-modal Retrieval-augmented Encoder}
\label{sec:cre}
To efficiently integrate intra-sample and inter-sample reference context into the target modality, we design the cross-modal retrieval-augmented encoder (CRE). It consists of an intra-sample cross-modal augmented encoder (Inter-CAE) and an inter-sample cross-modal augmented encoder (Intra-CAE).  The target modality and reference context are first passed through the self-attention \cite{vaswani2017attention} layer of Inter-CAE and Intra-CAE.  Then the intra-sample cross-modal augmented encoder uses modality-level reference contexts to enhance the target modality, while the inter-sample cross-modal augmented encoder uses sample-level reference contexts to reinforce the target modality. Formally,
\begin{equation}
\begin{aligned}
    &\hat{x}^{*_m} = \text{Inter-CAE}(x^{*}, x^{*_m}),*\in\{t,v,a\}\\
               &= softmax(\frac{(w^{inter}_{q_*} x^{*})({x^{*_m}}^{T}{w^{inter}_{k_*}}^{T})}{\sqrt{d{}}})(w^{inter}_{v_*} x^{*_m})) + x^{*}
\end{aligned}
\end{equation}

\begin{equation}
\begin{aligned}
    & \hat{x}^{*_s} = \text{Intra-CRE}(x^{*}, x^{*_s}),*\in\{t,v,a\}\\
               &= softmax(\frac{(w^{intra}_{q_*} x^{*})({x^{*_s}}^{T}{w^{intra}_{k_*}}^{T})}{\sqrt{d{}}})(w^{intra}_{v_*} x^{*_s})) + x^{*}
\end{aligned}
\end{equation}
\begin{equation}
     x_{final} = \text{Multimodal Fusion}(\hat{x}^{*_m}, \hat{x}^{*_s})
\end{equation}
 where $\hat{x}^{*_m}$ and $\hat{x}^{*_s}$ represent the enhanced target multimodal features that leverage modality-level reference context $x^{*_m}$ and sample-level reference context $x^{*_s}$, respectively.  The $w^{\{inter,inter\}}_{\{q_*,k_*,v_*\}}$ denotes parameter matrix. LN and FFN represent layer normalization and feed-forward network and are used to handle enhanced target multimodal features. Finally, these enhanced target multimodal features are fused and passed through several linear layers and activation functions for sentiment polarity prediction.

\subsection{Optimization Objective}
The optimization objective consists of two parts: the multimodal sentiment analysis task loss and the contrastive cross-modal retrieval loss. We use MSE Loss $\mathcal{L}_{msa}$ for the multimodal sentiment analysis task. The total loss is calculated as follows.
\begin{equation}
    \mathcal{L}_{total} = \mathcal{L}_{msa} + \lambda\mathcal{L}_{ccrl}
\end{equation}
where $\lambda$ denotes the hyper-parameter.

\begin{table*}[t]
    \caption{Performance comparison between our proposed TC$^2$RAHP and several state-of-art baselines on CMU-MOSI and CMU-MOSEI. The left and right sides of '/' represent the mean and maximum values, respectively, of the five experiments. '-' represents a result that is not reported in the literature.
    }
    \centering
    \begin{tabular}{c|c|c|c|c|c|c|c|c|c}
    \toprule
    \multirow{2}{*}{Category} & \multirow{2}{*}{Models} & \multicolumn{4}{|c|}{CMU-MOSI} & \multicolumn{4}{|c}{CMU-MOSEI} \\
    
    \cmidrule{3-6} \cmidrule{7-10}
    & & Acc2 $\uparrow$ & F1 $\uparrow$ & Corr $\uparrow$ & MAE $\downarrow$ & Acc2 $\uparrow$ & F1 $\uparrow$  & Corr $\uparrow$ & MAE $\downarrow$ \\
    \midrule 
    \multirow{11}{*}{MSA Model}
    &TFN & -/80.8 & -/80.7  & 0.698 & 0.901 & -/82.5 & -/82.1 & 0.700 & 0.593 \\

    &LMF & -/82.5 & -/82.4  & 0.695 & 0.917 & -/82.0 & -/82.1  & 0.677 & 0.623 \\

    &Mult & 81.5/84.1 & 80.6/83.9 & 0.711 & 0.861 & -/82.5 & -/82.3 & 0.703 & 0.580 \\

    &MISA & 80.79/82.10 & 80.77/82.03 & 0.764 & 0.804 & 82.59/84.23 & 82.67/83.97 & 0.724 & 0.568 \\

    &Self-MM & 82.54/84.77 & 82.68/84.91 & 0.795 & 0.712 & 82.68/84.96 & 82.95/84.93 & 0.767 & 0.529 \\

    &MMIM & 84.14/86.06 & 84.00/85.98 & 0.800 & 0.700 & 82.24/85.97 & 82.66/85.94 & 0.772 & \textbf{0.526} \\

    &FDMER & -/84.6 & -/84.7 & 0.788 & 0.724 & -/86.1 & -/85.8  & 0.773 & 0.536 \\

    &AMML & -/84.9 & -/84.8  & 0.792 & 0.723 & -/85.3 & -/85.2 & 0.776 & 0.614 \\

    &ConKI & 84.37/86.13 & 84.33/86.13  & 0.816 & 0.681 & 82.73/86.25 & 83.08/86.15 & 0.782 & 0.529 \\

    &HyDiscGAN & 84.1/86.7 & 83.7/86.3  & 0.782 & 0.749 & 81.9/86.3 & 82.1/86.2 & 0.761 & 0.533 \\

    &MOFN & 84.84/86.89 & 84.75/86.86  & 0.797 & 0.725 & 82.70/86.32 & 83.13/86.29  & 0.780 & 0.528 \\
    
    &DMD & -/86.00 & -/86.00  & - & - & -/86.60 & -/86.60  & - & - \\
    
    &ConFEDE & 84.17/85.52 & 84.13/85.52  & 0.784 & 0.742 & 81.65/85.82 & 82.17/85.83  & 0.780 & 0.522 \\
    &DTN & -/86.20 & -/86.20  & 0.807 & 0.714 & -/86.30 & -/86.30  & 0.788 & 0.579 \\
    \midrule
    \multirow{10}{*}{MLLM and LLM}
    &GPT-4V & -/90.91 & - & - & - & -/87.10 & -  & - & - \\
    &Claude3-V & -/78.79 & - & - & - & -/79.93 & -  & - & - \\
    &Gemini-V & -/88.34 & - & - & - & -/87.14 & -  & - & - \\
    &Chatgpt & -/89.60 & - & - & - & -/84.97 & -  & - & - \\
    &LLama2-7B & -/67.68 & - & - & - & -/77.30 & -  & - & - \\
    &LLama2-13B & -/81.86 & - & - & - & -/81.66 & - & - & - \\
    &Flan-T5-XXL & -/89.60 & - & - & - & -/86.52 & -  & - & - \\
    &QwenVL-Chat & -/85.93 & - & - & - & -/80.41 & -  & - & - \\ 
    &BLIP-2 & -/88.99 & - & - & - & -/86.88 & -  & - & - \\    
    &Instruct-BLIP & -/88.68 & - & - & - & -/85.98 & -  & - & - \\   
    \midrule
    Our Model& \textbf{TC$^2$RAHP} & \textbf{90.55/91.31} & \textbf{90.52/91.27}  & \textbf{0.873} & \textbf{0.630} & \textbf{87.10/87.58} & \textbf{87.12/87.39}  & \textbf{0.831} & 0.567 \\  
    \bottomrule
    \end{tabular}
    \label{tab:my_label}
\end{table*}

\section{EXPERIMENTS}
\subsection{Datasets}
In this paper, following previous work, we conducted experiments on two datasets: CMU-MOSI and CMU-MOSEI, to evaluate the performance of our model.

\textbf{CMU-MOSI} \cite{zadeh2016mosi}. 
The CMU-MOSI dataset serves as a widely recognized benchmark for sentiment analysis across multiple modalities, including acoustic, text, and visual. It consists of 2,199 YouTube video clips, each of which has been manually labeled with a sentiment intensity rating on a scale from -3 (indicating extreme negativity) to +3 (indicating extreme positivity).

\textbf{CMU-MOSEI} \cite{zadeh2018multimodal}.
CMU-MOSEI is a sentiment analysis dataset collected from YouTube. As an extension of CMU-MOSI, CMU-MOSEI shares a similar annotation standard for sentiment scores and encompasses a more diverse range of video categories, comprising a total of 22,856 video segments.
\subsection{Experimental Settings}
\paragraph{Evaluation Protocols.} 
Following previous work, we employed four evaluation metrics to assess the model's performance: Binary Classification Accuracy (Acc2), F1-Score, Mean Absolute Error (MAE), and Pearson Correlation (Corr). Specifically, we converted the regression predictions into binary values for computing Acc-2. Higher values indicate better performance for all metrics except MAE.
\paragraph{Implementation Details.}
We trained our proposed model using the PyTorch library on a single NVIDIA A100 GPU. The AdamW optimizer is configured with a batch size of 8. For training on the CMU-MOSI dataset, the number of epochs was set to 50, while for the CMU-MOSEI dataset, it was set to 20. The learning rate was fixed at 1e-5. The upper bound of the distance metric $\gamma$ was set to 50. 

\subsection{Baselines}
\paragraph{MSA Models.}
We compared our proposed model with state-of-the-art baseline models in the task of multimodal sentiment analysis to justify its effectiveness. \textbf{TFN} \cite{zadeh2017tensor} expands at the tensor level to capture both intra-modal and inter-modal information. \textbf{LMF} \cite{liu2018efficient} decomposes the model weights into low-rank factors for tensor fusion. \textbf{MULT} \cite{tsai2019multimodal} leverages cross-modal attention to incorporate information from other modalities into the current modality. \textbf{MISA} \cite{hazarika2020misa} divides the modality features into more fine-grained shared and exclusive features. \textbf{Self-MM} \cite{yu2021learning} generates labels through a self-supervised approach and further employs them to facilitate the interaction of multimodal information. \textbf{MMIM} \cite{han2021improving} captures sentiment information through multidimensional interactions between modalities. \textbf{FDMER} \cite{yang2022disentangled} optimizes the parameters of the encoders for identifying private and shared features through adversarial learning. \textbf{AMML} \cite{sun2022learning} integrates the information learned from single-modal tasks into the multimodal context. \textbf{ConKI} \cite{yu2023conki} incorporates modality-specific information for each modality. \textbf{HyDiscGAN} \cite{wu2024hydiscgan} aligns different modalities and generates simulated multimodal features from the obtained information to achieve privacy protection. \textbf{MOFN} \cite{chen2022mofn} proposes an alignment method that enriches the representation of video frame features to achieve video restoration. \textbf{DMD} \cite{li2023decoupled} addresses the modality imbalance issue using a cross-modal distillation method. \textbf{ConFEDE} \cite{yang2023confede} decomposes the features and enhances the information representation through contrastive learning. \textbf{DTN} \cite{zeng2024disentanglement} decouples the features of a single modality and explores the commonality and diversity between different modalities.
\paragraph{LLMs and MLLMs.} We selected several high-performing large language models from the benchmark tests \cite{yang2024mm} of multimodal large language models as baselines, such as \textbf{Chatgpt} \cite{chatgpt2023}, \textbf{LLama2} \cite{touvron2023llama}, \textbf{Flan-T5} \cite{chung2024scaling}, \textbf{Qwen-VLs} \cite{bai2023qwen}, \textbf{BLIP-2} \cite{li2023blip2bootstrappinglanguageimagepretraining}, \textbf{InstructBLIP} \cite{dai2023instructblipgeneralpurposevisionlanguagemodels}, \textbf{GPT-4V} \cite{gpt4v}, \textbf{Claude3-V} \cite{claude}, and \textbf{Gemini-V} \cite{geminiv}. 

\begin{figure}
    \centering
    \includegraphics[width=0.50\textwidth]{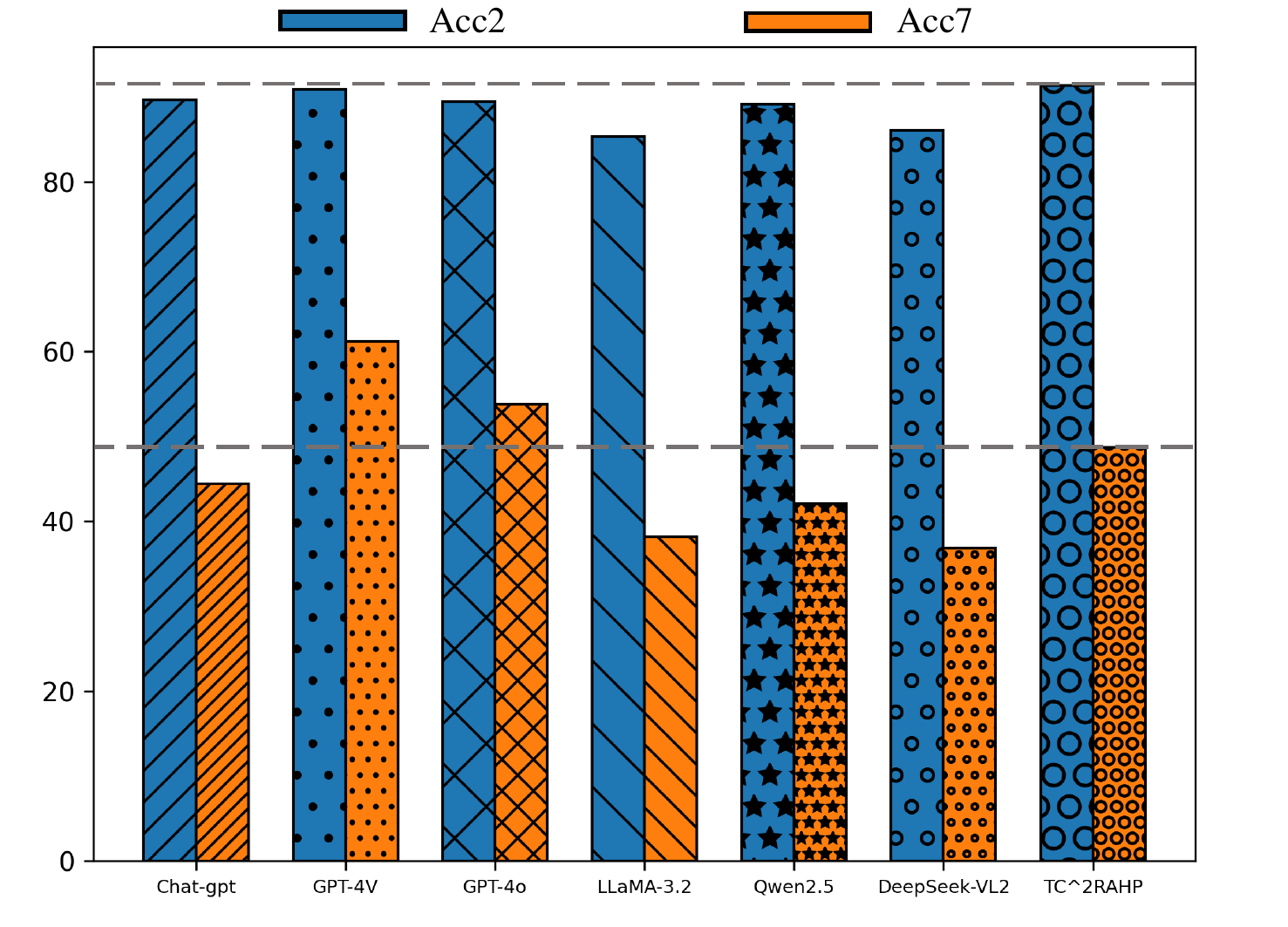}
    \caption{The performance of our proposed model with several state-of-the-art large models on the MOSI dataset. The specific versions of LLaMA-3.2 and Qwen2.5 are LLaMA-3.2-11B-Vision and Qwen2.5-VL-7B-Instruct.
    }
    \label{fig:acc7-mosi}
\end{figure}
\begin{figure}
    \centering
    \includegraphics[width=0.50\textwidth]{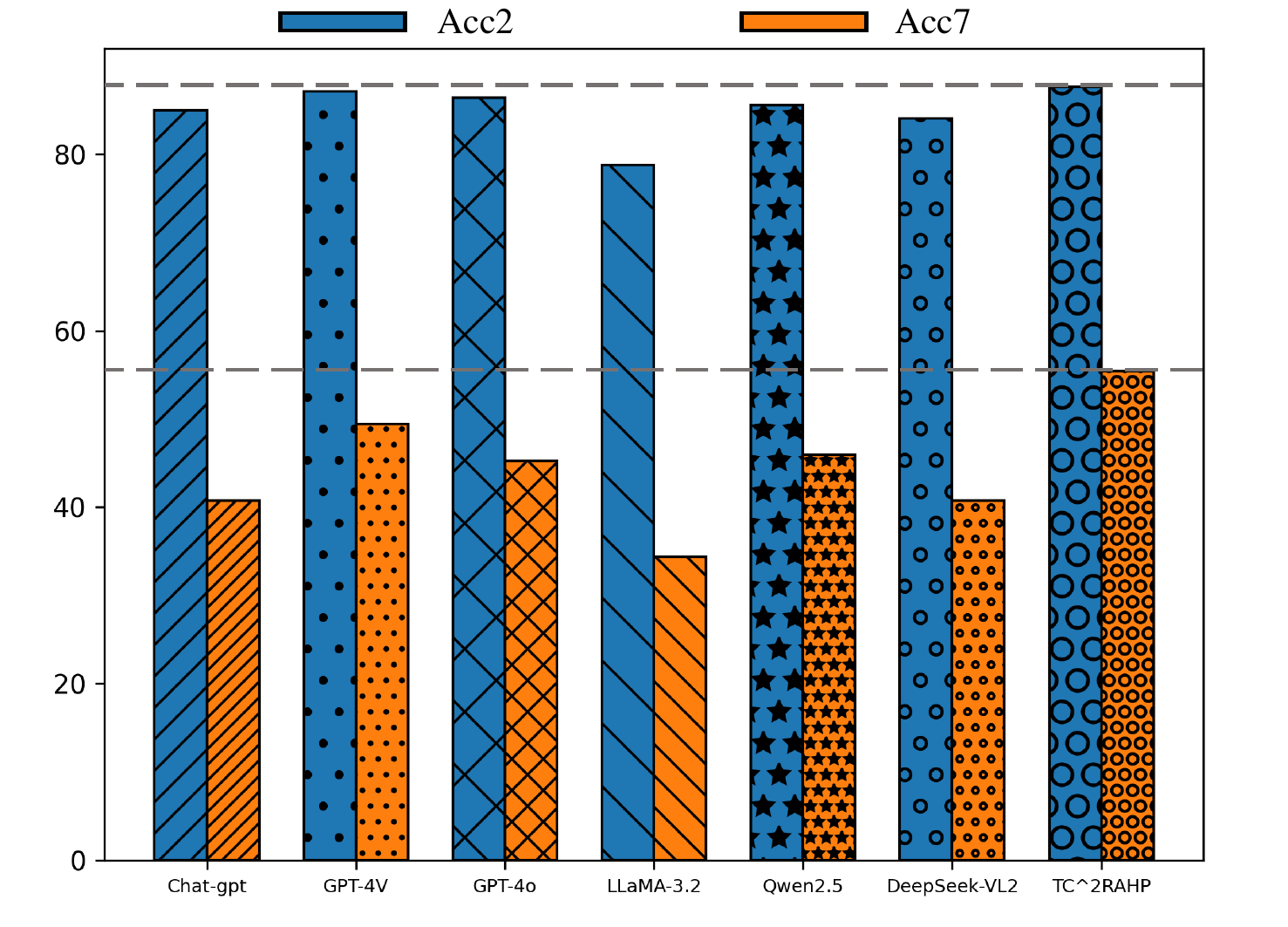}
    \caption{The performance of our proposed model with several state-of-the-art large models on the MOSEI dataset. The specific versions of LLaMA-3.2 and Qwen2.5 are LLaMA-3.2-11B-Vision and Qwen2.5-VL-7B-Instruct.
    }
    \label{fig:acc7-mosei}
\end{figure}

\subsection{Performance Comparison}
The performance comparisons are summarized in Table \ref{tab:my_label}. By analyzing this table, we gained the following observations: 1) Among traditional multimodal sentiment analysis methods, MOFN outperforms previous methods on both datasets by introducing a carefully designed compensation module to enhance video frames. Specifically, this compensation module aligns video frames, which are subsequently processed by a fusion module for reconstruction, thus obtaining frames of improved visual quality. This result indicates that fully leveraging the modality-level reference context is essential in sentiment analysis. 2) Among LLM and MLLM based methods, GPT-4V and Gemini-V outperforms the compared baselines. Compared with MOFN, GPT-4V and Gemini-V obtains relative Acc2 gains with 4.02\% and 0.82\%, respectively. This result indicates that LLMs possess representational capabilities that traditional sentiment analysis models cannot surpass.

Our proposed model outperforms all baseline models across all evaluation metrics on both the CMU-MOSI and CMU-MOSEI datasets. Compared with GPT-4V and Gemini-V, our proposed model obtains relative Acc2 gains with 0.4\% and 0.44\%, respectively. The reason is that our model not only focuses on the sample-level reference context but also incorporates the modality-level reference context. This indicates that simultaneously attending to both sample-level and modality-level reference contexts, and performing fine-grained interactions, can further enhance the model's  capability.

Beyond comparing the results under Acc-2, we also report performance under the Acc-7 metric, as shown in Figures \ref{fig:acc7-mosi} and \ref{fig:acc7-mosei}. The experimental results demonstrate that our model consistently outperforms all open-source large models and most closed-source models in terms of Acc-7. Notably, the closed-source models GPT-4v and GPT-4o were evaluated only on a subset of the dataset, which further highlights the effectiveness of our proposed model.

\begin{table}[t]
\setlength{\tabcolsep}{3mm}
    \caption{The ablation experiments on the CMU-MOSI and CMU-MOSEI datasets.}
    \centering
    \begin{tabular}{c|c|c|c|c}
    \toprule
    \multirow{2}{*}{Models} & \multicolumn{2}{c}{CMU-MOSI} & \multicolumn{2}{|c}{CMU-MOSEI} \\
    
    \cmidrule{2-3} \cmidrule{4-5}
    & Acc2 $\uparrow$ & F1 $\uparrow$  & Acc2 $\uparrow$ & F1 $\uparrow$   \\
    \midrule     

    w/o MMG  & 89.63 & 89.63  & 86.52 & 86.38 \\

    w/o SMG & 89.79 & 89.77  & 86.77 & 86.50 \\

    w/o (M)ICAE & 90.55 & 90.53   & 87.13 & 86.89  \\

    w/o (S)ICAE  & 90.85 & 90.82   & 87.19 & 87.23  \\

    \midrule
    \textbf{TC$^2$RAHP} & \textbf{91.31} & \textbf{91.27}  & \textbf{87.58} & \textbf{87.39}   \\  
    \bottomrule
    \end{tabular}
    \label{tab:my_label2}
\end{table}

\subsection{Ablation Study}
To gain a deeper understanding of our model's performance, we conducted ablation tests according to the following scheme. 1) w/o MMG, eliminating the Modality-level modality Generation module; 2) w/o SMG, excluding the Sample-level Modality Generation moudle. 3) w/o (M)ICAE, removing the Intra-sample Cross-modal Augmented Encoder, which is previously connected to Modality-level modality Generation module; 4) w/o  (S)ICAE, excluding the Inter-sample Cross-modal Augmented Encoder, which is previously connected to Sample-level Modality Generation module.
The experimental results are shown in Table \ref{tab:my_label2}.

Compared with our model, w/o MMG causes a greater performance drop than w/o SMG. The reason is that the modality-level reference context shares the same semantic space as the target sample, which reveals that the more similar the semantic spaces are, the more useful the information obtained through interaction, thereby better enhancing the target sample. Moreover, the performance drop of w/o (M)ICAE and w/o (S)ICAE can be observed, revealing that enabling the target sample to interact with both the modality-level and sample-level reference contexts effectively enhances the feature representations.  

\begin{figure}
    \centering
    \includegraphics[width=0.5 \textwidth]{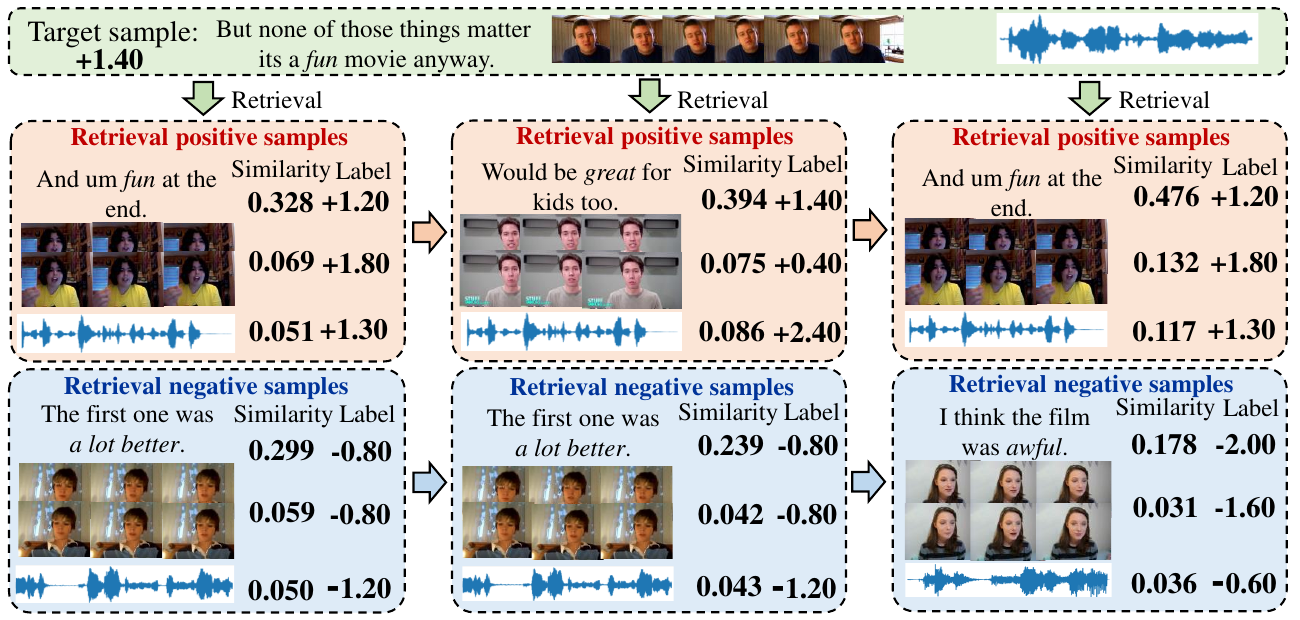}
    \caption{
    Visualization of the changes in the positive and negative samples retrieved for a specific target modality during the training process.
    }
    \label{fig:retrieval}
\end{figure}

\subsection{Cross-modal Retrieval Sample Visualization}
Apart from achieving outstanding performance, the key advantage of TC$^2$RAHP is that it can introduce the sample-level reference context to enhance target sample through contrastive cross-modal retrieval module. To this end, we visualized the process of retrieving positive and negative samples for the target modality during the training phase. 
Specifically, we selected specific samples, using the text modality as an example, and visualized the changes in the retrieved positive and negative samples as the training steps increased, along with the corresponding similarity values. The results are displayed in Figure \ref{fig:retrieval}. From Figure \ref{fig:retrieval}, we gained the following observations. 1) In the initial stage, the similarity of positive and negative samples relative to the target modality is similar, but their semantic information is opposite. 2) As the training steps increase, the similarity of positive samples to the target modality increases, while the similarity of negative samples to the target modality decreases. 3) The retrieved positive and negative samples are dynamically changing. This result indicates that relying solely on similarity-based retrieval may lead to samples with opposite semantics, which could negatively impact the model. After adding the contrastive cross-modal retrieval constraint, the similarity of samples with opposing semantics to the target modality is significantly reduced. This strongly demonstrates the effectiveness of the contrastive cross-modal retrieval module.

\begin{figure}
    \centering
    \includegraphics[width=0.5 \textwidth]{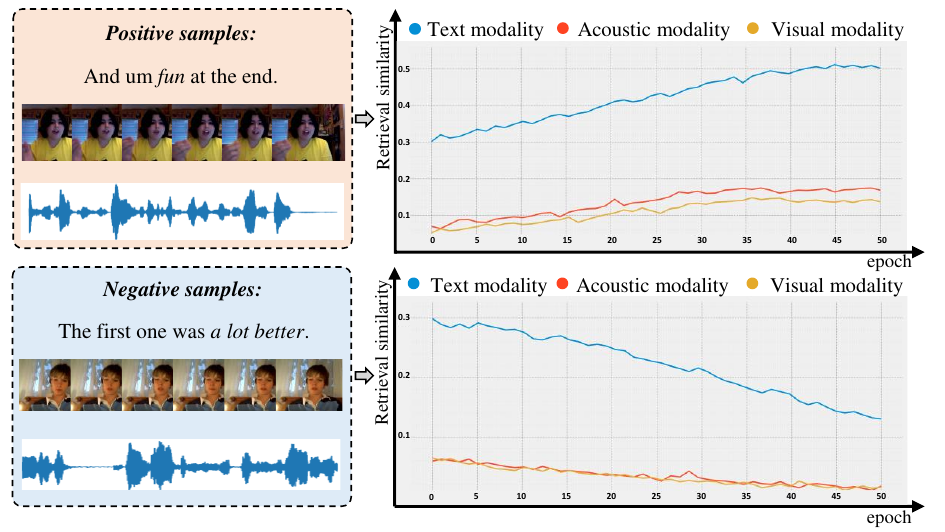}
    \caption{The evolution of similarity between the target sample and the corresponding positive and negative samples during training stage. 
    }
    \label{fig:sim}
\end{figure}

\begin{figure}
    \centering
    \includegraphics[width=0.50\textwidth]{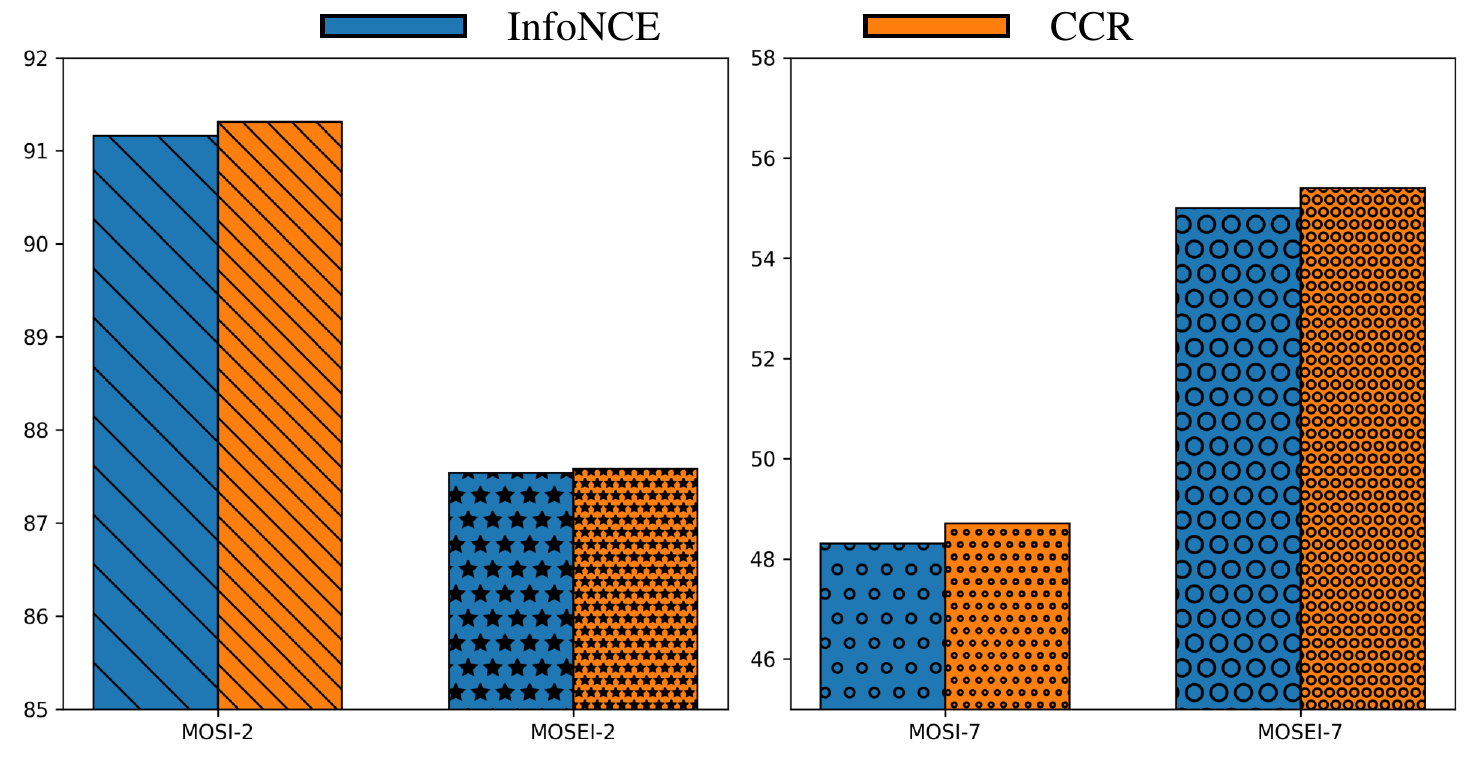}
    \caption{Comparison between our proposed  $\mathcal{L}_{ccr}$ and InfoNCE loss on the MOSI and MOSEI datasets.
    }
    \label{fig:ccr-infoNce}
\end{figure}

To gain deeper insights into the contrastive cross-modal retrieval module, we explained the changes in the similarity of specific target modality and the retrieved positive and negative samples as the training steps increase, as shown in Figure \ref{fig:sim}. We could observe that the similarity between the target modality and the fixed retrieved positive samples increases, while the similarity with the negative samples exhibits an opposite trend. This result further confirms the effectiveness of the contrastive cross-modal retrieval module. This result further demonstrates the effectiveness of the contrastive cross-modal retrieval module, as it ensures that the model retrieves samples that are similar to the target modality both in terms of features and semantics, thereby enhancing the model's performance. In addition to visualizing the retrieval process, we also compare different contrastive learning strategies. As shown in Figure \ref{fig:ccr-infoNce}, our proposed $\mathcal{L}_{ccr}$ achieves comparable performance to the widely used InfoNCE loss, with a slight advantage.

\begin{figure}
    \centering
    \includegraphics[width=0.50\textwidth]{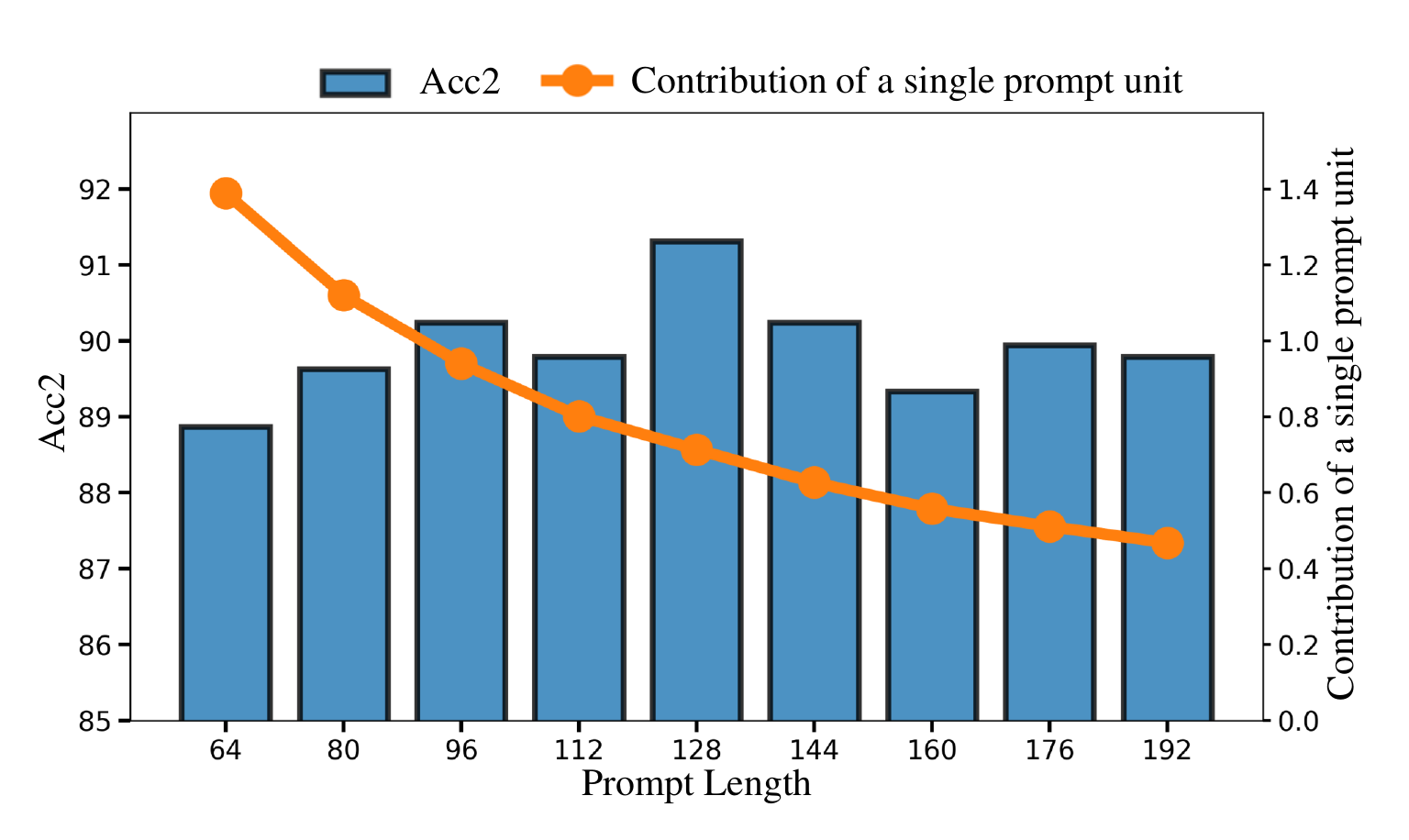}
    \caption{The analysis of the impact of prompt length on model performance on the CMU-MOSI dataset. The left and right y-axes represent accuracy and the contribution of prompt units to the model's performance, respectively.
    The x-axis represents the prompt length.
    }
    \label{fig:prompt}
\end{figure}
\subsection{Hierarchical Prompt Analysis}
To validate the effectiveness of hierarchical prompt, we conducted experiments to observe the impact of prompt length on model performance. The results are shown in Figure \ref{fig:prompt}. The left and right y-axes represent binary accuracy and the contribution rate of prompt units, respectively. The prompt unit contribution rate measures the contribution of prompt units of unit length to the model's performance. From Figure \ref{fig:prompt}, we could observe that the model's performance increases initially with the length of the prompt, then slightly decreases. The unit prompt contribution rate is decreasing. The model achieves the best performance when the prompt length is 128. This may be due to the fact that as the prompt length increases, the model gradually learns more information from both modality-level and sample-level reference contexts. The shorter prompt length fails to carry enough reference context information, while the longer prompt length causes the model to focus on trivial details, leading to a decline in performance. This result proves the effectiveness of modality-level and sample-level prompts, as well as the fact that an appropriate prompt length can enhance model performance.

\begin{figure}
    \centering
    \includegraphics[width=0.5 \textwidth]{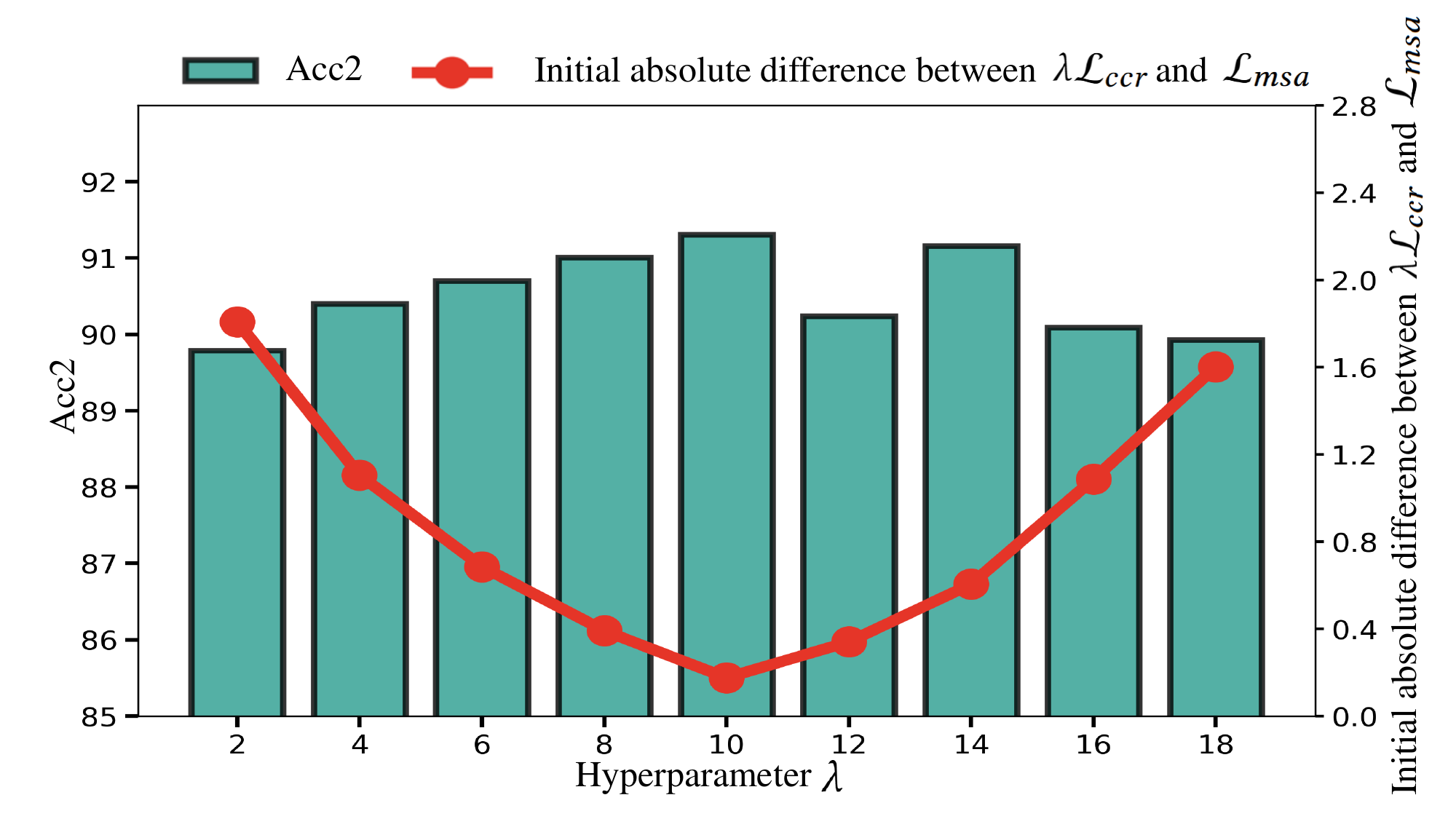}
    \caption{The experiment analyzing the impact of hyperparameter $\lambda$ on model performance conducted on the CMU-MOSI dataset. The x-axis corresponds to hyperparameter $\lambda$ ($\times 10^{-4}$), and the y-axes corresponds to Acc2 and the initial absolute difference between $\lambda\mathcal{L}_{ccr}$ and $\mathcal{L}_{msa}$.}
    \label{fig:alpha}
\end{figure}

\subsection{Hyper-parameter analysis}
To explore the impact of the hyperparameter $\lambda$, we conducted experiments by increasing it from 0.0002 to 0.0018. The left y-axis corresponds to Acc2 and the right y-axis corresponds to the initial absolute difference between $\lambda\mathcal{L}_{ccr}$ and $\mathcal{L}_{msa}$. Figure \ref{fig:alpha} displayed the experimental results. We could observe that the absolute difference between the two task losses first decreases and then increases, while the model performance initially improves and then declines. This can be attributed to the significant gap in task losses, indicating that the sub-tasks were not fully optimized. It also highlights the importance of the sub-task design, particularly the optimization of the contrastive cross-modal retrieval loss.

\begin{figure}
    \centering
    \includegraphics[width=0.5 \textwidth]{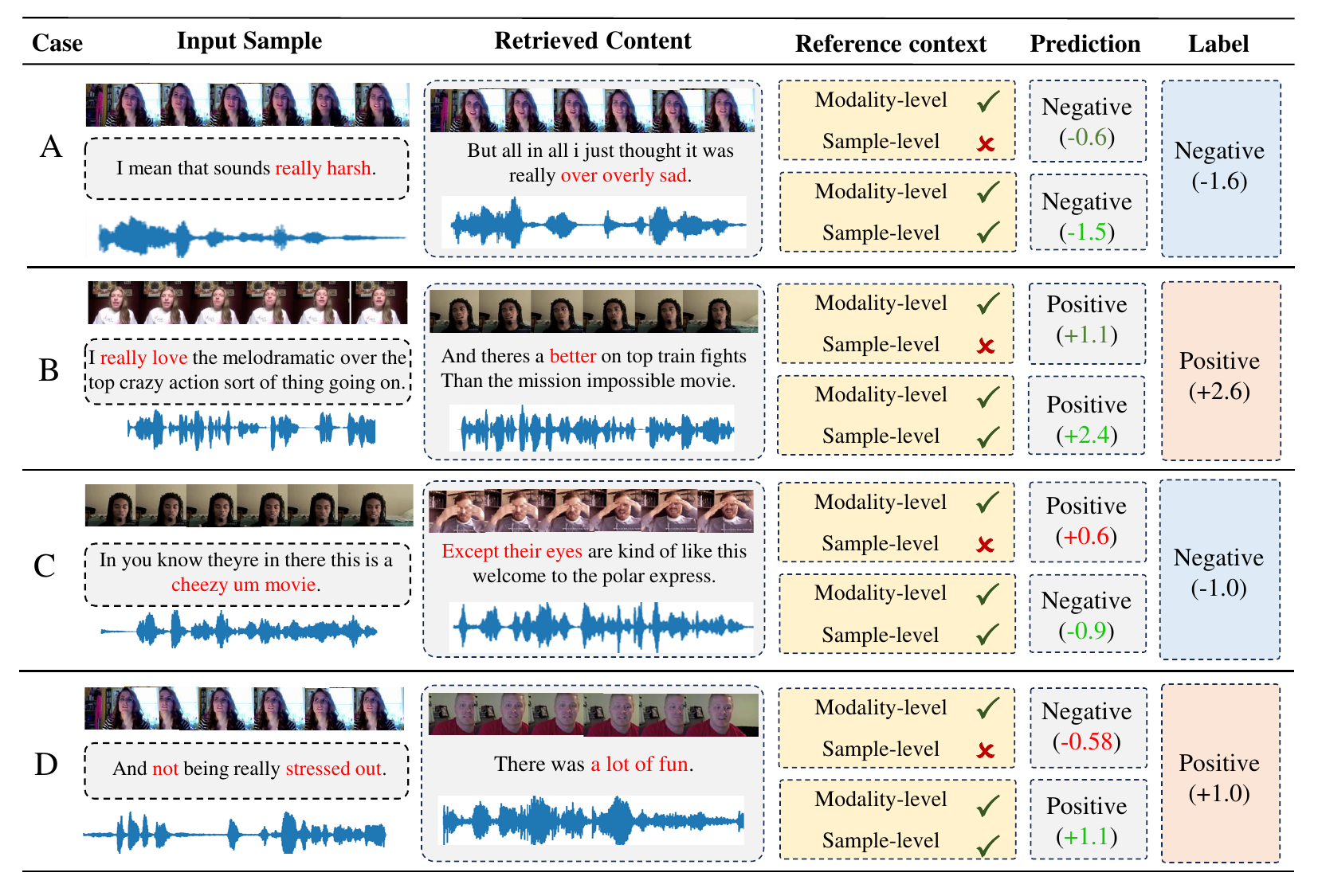}
    \caption{Case study on CMU-MOSI dataset. The model predicts values using only modality-level reference context to enhance the target modality, and compares this with predictions made by simultaneously using modality-level context and sample-level context to enhance the target modality.}
    \label{fig:case}
\end{figure}
\subsection{Case Study}
To qualitatively validate the effectiveness of our proposed model, we displayed several typical results, as shown in Figure \ref{fig:case}, which are made using only modality-level reference context and those made using both modality-level reference context and sample-level reference context. Based on these experimental results, we could observe that solely utilizing sample-level reference context to enhance the modality may lead to inaccurate or even incorrect sentiment polarity. The additional incorporation of sample-level reference context guides the model to output the correct sentiment polarity. This qualitative experimental result highlights the necessity and effectiveness of introducing sample-level reference context.

\section{Conclusion}
In this paper, we present a multimodal retrieval-enhanced framework for multimodal sentiment analysis tasks. It is the first work to explore the direct integration of external samples to enhance modalities through the retrieval mechanism. Specifically, we first design a cross-modal contrastive retrieval module that retrieves samples semantically similar to the target modality. We then introduce modality-level and sample-level prompts, leveraging inter-sample and intra-sample information to generate modality-level and sample-level reference contexts, respectively. Finally, we design a cross-modal retrieval-enhanced encoder that utilizes modality-level and sample-level reference contexts to enhance the target modality, improving its representational capacity. Extensive experiments demonstrate the effectiveness and superiority of the proposed model.

\bibliographystyle{IEEEtran}
\bibliography{mybib.bib}


 





\end{document}